\documentclass[twocolumn,prb,superscriptaddress]{revtex4-2}

\usepackage{amsmath,amsfonts,amssymb}
\usepackage{graphicx,color,xcolor}
\usepackage{hyperref}
\usepackage{epstopdf}
\usepackage{float}
\usepackage{multirow}
\usepackage{hhline}
\usepackage{ulem}
\usepackage{mathtools}
\usepackage{siunitx}
\usepackage[caption=false]{subfig}
\usepackage{soul}
\hypersetup{hidelinks,colorlinks=true,allcolors=BLUE}

\newcommand{\ket}[1]{\left|#1\right\rangle}
\newcommand{\bra}[1]{\left\langle#1\right|}

\newcommand{\BT}[1]{\textcolor{blue}{ #1}}

\makeatletter

\@ifundefined{textcolor}{}
{%
 \definecolor{BLACK}{gray}{0}
 \definecolor{WHITE}{gray}{1}
 \definecolor{RED}{rgb}{1,0,0}
 \definecolor{GREEN}{rgb}{0,1,0}
 \definecolor{BLUE}{rgb}{0,0,1}
 \definecolor{CYAN}{cmyk}{1,0,0,0}
 \definecolor{MAGENTA}{cmyk}{0,1,0,0}
 \definecolor{YELLOW}{cmyk}{0,0,1,0}
}

\makeatother

\begin{document}

\title{Hardware efficient autonomous error correction with linear couplers in superconducting circuits}

\author{Ziqian Li$^\ddag$}
\affiliation{James Franck Institute, University of Chicago, Chicago, Illinois 60637, USA}
\affiliation{Department of Physics, University of Chicago, Chicago, Illinois 60637, USA}
\affiliation{Department of Applied Physics, Stanford University, Stanford, California 94305, USA}
\altaffiliation{These authors contributed equally to this work}

\author{Tanay Roy$^\ddag$}
\email{Present address: Superconducting Quantum Materials and Systems Center, Fermi National Accelerator Laboratory (FNAL), Batavia, IL 60510, USA}
\affiliation{James Franck Institute, University of Chicago, Chicago, Illinois 60637, USA}
\affiliation{Department of Physics, University of Chicago, Chicago, Illinois 60637, USA}
\altaffiliation{These authors contributed equally to this work}

\author{David Rodr{\'i}guez P{\'e}rez}
\affiliation{Department of Physics, Colorado School of Mines, Golden, Colorado 80401}

\author{David I. Schuster}
\affiliation{James Franck Institute, University of Chicago, Chicago, Illinois 60637, USA}
\affiliation{Department of Physics, University of Chicago, Chicago, Illinois 60637, USA}
\affiliation{Pritzker School of Molecular Engineering, University of Chicago, Chicago, Illinois 60637, USA}
\affiliation{Department of Applied Physics, Stanford University, Stanford, California 94305, USA}

\author{Eliot Kapit}
\affiliation{Department of Physics, Colorado School of Mines, Golden, Colorado 80401}

\date{\today}

\begin{abstract}

Large-scale quantum computers will inevitably need quantum error correction (QEC) to protect information against decoherence. Given that the overhead of such error correction is often formidable, autonomous quantum error correction (AQEC) proposals offer a promising near-term alternative. AQEC schemes work by transforming error states into excitations that can be efficiently removed through engineered dissipation. {The recently proposed AQEC scheme by Li et al., called the Star code, can autonomously correct or suppress all single qubit error channels using two transmons as encoders with a tunable coupler and two lossy resonators as a cooling source.} The Star code requires only two-photon interactions and can be realized with linear coupling elements, avoiding experimentally challenging higher-order terms needed in many other AQEC proposals,  but needs carefully selected parameters to achieve quadratic improvements in logical states' lifetimes. Here, we theoretically and numerically demonstrate the optimal parameter choices in the Star Code. We further discuss adapting the Star code to other planar superconducting circuits, which offers a scalable alternative to single qubits for incorporation in larger quantum computers or error correction codes.
\end{abstract}

\maketitle

\section{Introduction}
Random interactions with the environment accumulate errors in qubits, strongly degrading the performance of modern quantum computers. Fundamentally, random errors are a source of classical entropy that heats the system away from its target states, and a cooling mechanism--quantum error correction--is required to continuously remove this entropy. This cooling is particularly important for large-scale processors where information needs to be stored for longer periods and shuffled across longer distances. Further, the correction must be done without learning the state of the qubits. It is widely believed that fault tolerance will ultimately require topological codes such as the surface code \cite{fowler2012surface,terhal2015}, where the quantum information is encoded in the collective state of a topological field theory and error correction is performed by repeated measurement and feedback. 

However, the precision control requirements, overhead in qubit count, wall clock time \cite{babbush2021focus} and classical processing in fault-tolerant codes pose significant challenges for practical implementation. A compelling complementary solution, both for scaling near-term algorithms and as a component qubit in larger fault tolerant codes, is autonomous QEC (AQEC) \cite{wang2022, Albert_2019, eliot2018, Lihm2018, eliot2016, Joachim2014, Mirrahimi_2014, Zaki2013, Mohan2005, kapit2017review, kapit2022small, Gertler2021, Grimm2020, Campagne-Ibarcq2020, Ma2020, Hu2019, Ofek2016}. AQEC methods are based on carefully engineering the level structure and external drives applied to small clusters of component devices, such as transmon qubits and resonators, so that states created by errors are assigned an energy penalty and can be rapidly corrected using engineered dissipation. But of necessity, these devices can still be fairly complex, and in the case of 3D cavity proposals, physically large. Finding compact, efficient implementations of AQEC is thus an important area of research.

A compelling early AQEC proposal is the Very Small Logical Qubit (VSLQ) architecture \cite{eliot2016,eliot2018,perez2020improved}. The VSLQ encodes a logical state using just two transmon qubits, each using the lowest three levels to encode the information. The VSLQ is able to achieve significant reductions in both idle and gate error (via error transparency and/or divisibility \cite{eliot2018,perez2021error}) by exploiting the empirical structure of noise in superconducting circuits, which is massively dominated by photon loss and low-frequency phase noise. The VSLQ uses continuous four-photon drive terms and blue sideband couplings to lossy resonators to autonomously correct photon loss, automatically suppressing phase noise in the process by generating a substantial energy penalty for local $Z$ operations. However, the high-order nonlinear terms in its Hamiltonian are very difficult to implement in practice, requiring unusual circuit elements and very high-frequency drives, thus far preventing its complete realization. Note that the requirement of four-photon drive and/or dissipative processes to stabilize the codewords is found in autonomous implementations of cat codes as well \cite{Zaki2013,Joachim2014,Mirrahimi_2014,mundhadagrimm2016,Ofek2016,albert2019pair,Gertler2021}.

{In this paper, we theoretically discuss the parameter choices for the recently proposed new AQEC code\cite{li2023autonomous}, called the Star code, to achieve quadratic lifetime improvement.} The Star code decomposes the four-photon processes of the VSLQ into simpler two-photon transitions (that can be engineered with a variety of AC-driven tunable coupler designs \cite{snail2017,luchakram2017, IBM2018, 2021rigetti, Brown2022, li2023autonomous}) and encodes the VSLQ codewords as a pair of degenerate dark states of the resulting rotating frame Hamiltonian. These simplifications make the Star code significantly easier to implement and allow it to be adapted to other types of a base qubit, such as fluxonia \cite{manucharyan2009fluxonium,xiong2022arbitrary,somoroff2021millisecond,zhang2021universal}, in a similarly compact, planar circuit. 

The structure of this paper is as follows. We begin by defining the Star code drive structure, encoding, and Hamiltonian. From it, we derive the resulting lifetime improvements analytically and compare them to numerical simulation. In doing so, we discuss each parameter's effect on code performance. We finally discuss gate protocols and extensions to other qubit types and offer concluding remarks.

\section{Star Code Protocol}

The Star code encodes a single logical qubit using the bottom three levels, $\ket{g}, \ket{e}$, and $\ket{f}$ of two transmons with the aim of correcting single-photon loss errors and suppressing dephasing in the process. The logical ``zero" and ``one" are defined as $\ket{L_0}=\left(\ket{gf}-\ket{fg}\right)/\sqrt{2}$ and $\ket{L_1}=\left(\ket{gg}-\ket{ff}\right)/\sqrt{2}$ respectively. Any valid quantum error correction codewords need to satisfy the Knill-Laflame conditions~\cite{KL2000}. With $a_{qj}$ representing the decay operator on the $j$-th transmon, these conditions can be stated as---(a) logical states are orthogonal, $\bra{L_1}{L_0}\rangle=0$; (b) error states are orthogonal, $\bra{L_1}a_{qj}^{\dagger}a_{qj}\ket{L_0}=0$; (c) error states are orthogonal with logical states, $\bra{L_1}a_{qj}\ket{L_0}=\bra{L_0}a_{qj}\ket{L_1}=\bra{L_0}a_{qj}\ket{L_0}=\bra{L_1}a_{qj}\ket{L_1}=0$; (d) the error does not distinguish logical states, $\bra{L_0}a_{qj}^{\dagger}a_{qj}\ket{L_0}=\bra{L_1}a_{qj}^{\dagger}a_{qj}\ket{L_1}$. Up to an irrelevant relative phase, these are the same codewords as in the original VSLQ proposal.

\begin{figure}[t]
    \centering
    \includegraphics[width=\columnwidth]{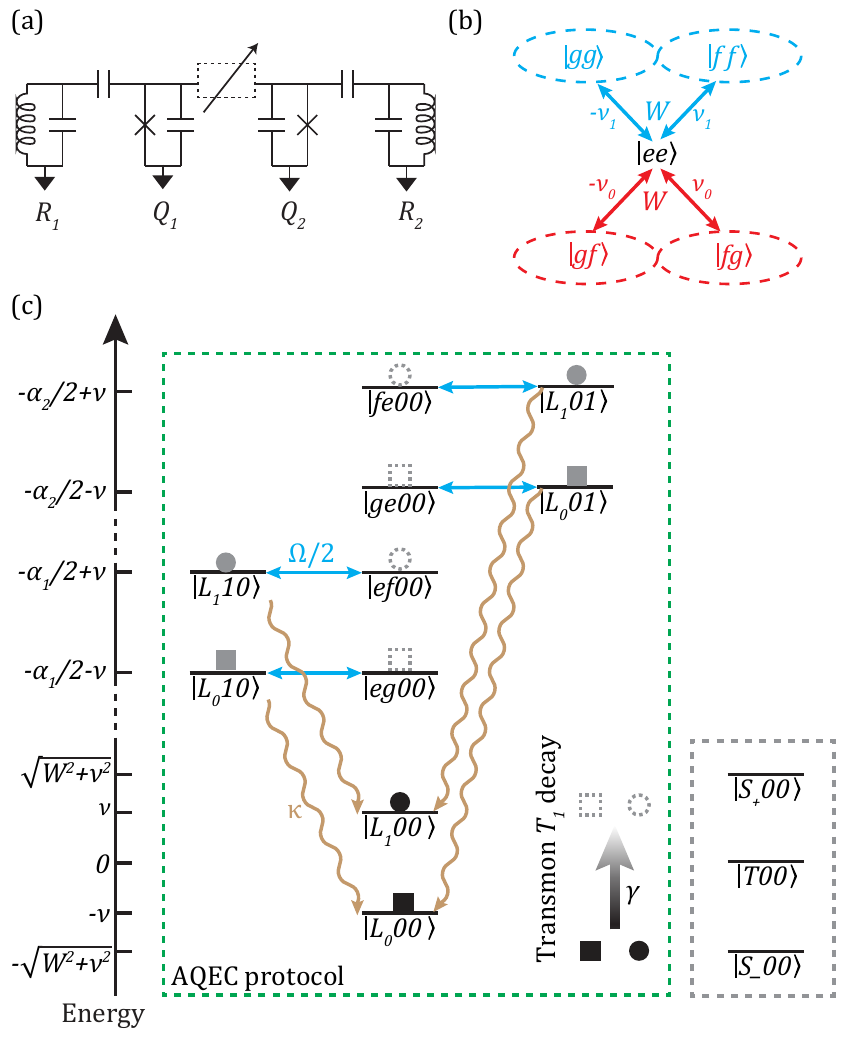}
    \caption{Star code protocol. (a) An example of hardware layout. Two transmons are individually coupled to two resonators dispersively. The dashed box between the two transmons represents any tunable coupling element that can provide sufficient QQ red and blue sideband interactions. (b) Four QQ sideband mixing configurations in the logical static frame. All sidebands are applied with an equal rate $W$ and specific detuning choices ($\pm \nu_{0/1}$) to construct the logical manifold. (c) Energy diagram in the rotating frame. The green dash box covers the logical states and error states involved in the AQEC protocol, and the grey dash box includes the other stray eigenstates that have suppressed population transfer by the energy gap. Error states from single-photon loss are restored to the parent logical states through individual correction paths. The QR sidebands (rate $\Omega/2$), the resonators' photon decay (rate $\kappa$), and the transmon $T_{1}$ decay (rate $\gamma$) are shown in the blue, brown, and black arrows respectively.}
    \centering
    \label{fig:energy_level}
\end{figure}

To implement AQEC, we need to engineer a continuously applied parent Hamiltonian, of which these two codewords are degenerate eigenstates. In Fig.~\ref{fig:energy_level}(a), we consider two transmons $Q_{1}$ and $Q_{2}$ with frequencies $\omega_{qj}$ and anharmonicities $\alpha_j$, which interact with each other through a tunable coupling element~\cite{li2023autonomous}. Two lossy resonators $R_{1}$ and $R_{2}$ dispersively coupled to the transmons have frequencies $\omega_{r1}$ and $\omega_{r2}$ separately, and we access the first two energy levels $\ket{0}$ and $\ket{1}$. The state for each transmon-resonator pair is labeled as $\ket{q,n}\in\{\ket{g},\ket{e},\ket{f}\}\otimes\{\ket{0},\ket{1}\}$. Assume the external drives can independently modulate the strength of the transversal interactions between two transmons (QQ) through the tunable coupling element and each transmon-resonator (QR) pair. The lab frame Hamiltonian of the full system is
\begin{align}
H_{\rm lab} =& \sum_{j=1}^{2}\left(\omega_{qj}a_{qj}^{\dagger}a_{qj} + \frac{\alpha_{j}}{2}a_{qj}^{\dagger}a_{qj}^{\dagger}a_{qj}a_{qj} + \omega_{rj}a_{rj}^{\dagger}a_{rj}\right) \nonumber\\
&+H_{QQ} + \sum_{j=1}^2 H_{QRj}, \label{eq:Hlab}\\
H_{QQ}=& A_{QQ}\left(t\right)\left(a_{q1}^{\dagger}+a_{q1}\right)\left(a_{q2}^{\dagger}+a_{q2}\right), \nonumber\\
H_{QRj}=& A_{QRj}\left(t\right)\left(a_{qj}^{\dagger}+a_{qj}\right)\left(a_{rj}^{\dagger}+a_{rj}\right), \nonumber\\
A_{QQ}\left(t\right) = & \frac{W}{\sqrt{2}}\cos{\left(\left(\omega_{q2}-\omega_{q1}-\alpha_{1}-\nu_{0}\right)t\right)} \nonumber\\
&+ \frac{W}{\sqrt{2}}\cos{\left(\left(\omega_{q2}-\omega_{q1}+\alpha_{2}+\nu_{0}\right)t\right)} \nonumber\\
&+ W\cos{\left(\left(\omega_{q1}+\omega_{q2}-\nu_{1}\right)t\right)} \nonumber\\
&+ \frac{W}{2}\cos{\left(\left(\omega_{q1}+\omega_{q2}+\alpha_{1}+\alpha_{2}+\nu_{1}\right)t\right)}, \nonumber\\
A_{QRj}\left(t\right) = & \frac{\Omega_{j}}{\sqrt{2}}\cos{\left(\left(\omega_{qj}+\omega_{rj}+\alpha_{j}\right)t\right)}.\nonumber 
\end{align}
In Eq.~\ref{eq:Hlab}, the QQ modulation is composed of four two-photon sidebands (two QQ red sidebands and two QQ blue sidebands) $\{\ket{ee}\leftrightarrow\ket{gf},\ket{ee}\leftrightarrow\ket{fg},\ket{ee}\leftrightarrow\ket{gg},\ket{ee}\leftrightarrow\ket{ff}\}$ with modulation amplitudes
$\{{W}/{\sqrt{2}},{W}/{\sqrt{2}},W, {W}/{2}\}$ and frequency detunings $\{\pm\nu_{0},\pm\nu_{1}\}$. The modulation amplitudes are chosen such that the oscillation rates between levels are the same for each sideband. The two QR sidebands generate the on-resonance transition $\ket{e0}\leftrightarrow\ket{f1}$ between transmon-resonator pairs. The QR modulation amplitudes $\Omega_{j}$ are kept small compared to $W$ in order to be treated as a perturbation to the system. 

We perform several rotating frame transformations while restricting the Hilbert space dimension to $3\times3\times2\times2$ ($Q_{1}Q_{2}R_{1}R_{2}$) for simplicity. We first define a series of transformation operators
\begin{align}
U_{1}(t)&=\exp\left[i\sum_{j=1,2}\left(\omega_{qj}+\frac{\alpha_{j}}{2}\right)a_{qj}^{\dagger}a_{qj}t\right],\\
U_{2}(t)&=\exp\left[-i\frac{\alpha_{1}+\alpha_{2}}{2}P_{ee}t\right],\\
U_{3}(t)&=\exp\left[i\nu_{0}(P_{gf}+P_{fg}+P_{ge}+P_{eg})t\right], \\
U_{4}(t)&=\exp\left[i\nu_{1}(P_{gg}+P_{ff}+P_{ef}+P_{fe})t\right], \\
U_{5}(t)&=\exp\left[i\sum_{j=1,2}(\omega_{rj}+\frac{\alpha_{j}}{2})a_{rj}^{\dagger}a_{rj}t\right].
\end{align}
Here we define $P_{ab}=\ket{ab}\bra{ab}\otimes I_{2}\otimes I_{2}$, where $I_{2}$ is a $2\times 2$ identity matrix.

There are two useful frames for intuitive understanding of the Star code with the transformation matrix $U_{a}=U_5U_2U_1$ (where all logical states are time-independent) and $U_b=U_5U_4U_3U_2U_1$ (where all interactions are time-independent). {In the first frame, the logical states are easy to write out directly, while the second frame is faster for simulation.} With $U_{a}$, the system Hamiltonian is transformed to the logical static frame $H_{\rm static}$. If the two frequency detunings $\nu_0$ and $\nu_1$ are unequal, there will be two time-independent zero-energy eigenstates $\{\ket{L_0},\ket{L_1}\}$ that form the static logical manifold. Applying rotating wave approximation (RWA) to $H_{QRj}$, the full system Hamiltonian becomes
\begin{align}
H_{\rm static} &= U_{a}H_{lab}U_{a}^{\dagger} + i\dot{U}_{a}U_{a}^{\dagger} \nonumber\\
&= \tilde{H}_{QQ} + \tilde{H}_{QR1} + \tilde{H}_{QR2} -\sum_{j=1,2}\frac{\alpha_{j}}{2}a_{rj}^{\dagger}a_{rj}\nonumber\\
&- \frac{\alpha_{1}}{2}(P_{eg}+P_{ef})-\frac{\alpha_{2}}{2}(P_{ge}+P_{fe}),\label{eq:static_frame}\\
\tilde{H}_{QQ} &= \frac{W}{2}\left[(\ket{gf}\bra{ee}+\ket{fg}\bra{ee})e^{-it\nu_{0}} \nonumber\right.\\
&\left.+(\ket{gg}\bra{ee}+\ket{ff}\bra{ee})e^{-it\nu_{1}}\right]+h.c.,\\
\tilde{H}_{QR1} &=\frac{\Omega_{1}}{2}\left(\ket{eg}\bra{fg}+\ket{ef}\bra{ff}\right)\otimes\ket{0}\bra{1}\otimes I_{2}+h.c., \label{eq:HQR1}\\
\tilde{H}_{QR2}&=\frac{\Omega_{2}}{2}\left(\ket{ge}\bra{gf}+\ket{fe}\bra{ff
}\right)\otimes I_{2}\otimes\ket{0}\bra{1}+h.c..
\end{align}

In the second frame where all interactions are time-independent, the system Hamiltonian $H_{\rm rot}$ is
\begin{align}
H_{\rm rot} &= U_{b}H_{lab}U_{b}^{\dagger} + i\dot{U}_{b}U_{b}^{\dagger}\nonumber\\
&= -\frac{\alpha_{1}}{2}(P_{eg}+P_{ef}) -\frac{\alpha_{2}}{2}(P_{ge}+P_{fe}) \nonumber\\
&-\nu_{0}(P_{gf}+P_{fg}+P_{ge}+P_{eg}) \nonumber\\
&-\nu_{1}(P_{gg}+P_{ff}+P_{ef}+P_{fe})\nonumber\\
&+H'_{QQ}-\sum_{j=1,2}(\frac{\alpha_{j}}{2}a_{rj}^{\dagger}a_{rj}+\tilde{H}_{QRj}),
\label{eq:fully_rotated}\\
H'_{QQ}&=\frac{W}{2}\left(\ket{ee}\bra{gf}+\ket{ee}\bra{fg}\right.\nonumber\\
&\left.+\ket{ee}\bra{gg}+\ket{ee}\bra{ff}+h.c.\right)\otimes I_{2}\otimes I_{2}. \label{eq:HQQ}
\end{align}

We assume the following hierarchy of rates for perturbation treatment of $\tilde{H}_{QRj}$: $W\gg\Omega_{j}\sim\kappa_{j}\gg\gamma_{j}$. Here $\{\Omega_{j}/2, \kappa_{j}, \gamma_{j}\}$ represent the QR$_{j}$'s sideband rate, R$_{j}$'s decay rate, and Q$_{j}$'s decay rate, respectively. Such hierarchies are generic features of AQEC schemes. For simplicity, in the following discussion, we assume $\Omega_{j}=\Omega$, $\kappa_{j}=\kappa$, and $-\nu_{1}=\nu_{0}=\nu$; error correction performance is not significantly impacted by small variations in $\Omega_j$ and $\kappa_j$ between the two component qubits.

In Fig.~\ref{fig:energy_level}(c), we plot the eigenstates for $H_{\rm rot}$ {in
the absence of the qutrit-resonator interaction} and sideband transitions between states to explain the AQEC process. The eigenstates can be grouped into three sets: $\{\ket{L_0}, \ket{L_1}\}$, $\{\ket{eg},\ket{ge},\ket{ef},\ket{fe}\}$, and $\{\ket{T},\ket{S_{-}},\ket{S_{+}}\}$. The first set forms the logical space with eigenenergies $\{-\nu,\nu\}$. The second set contains the states originating from a single photon loss error. The third set is comprised of stray eigenstates (not normalized for brevity) that are suppressed by the frequency detuning choice $\pm\nu$,
\begin{align}
\ket{T}&=\ket{gg}-\ket{gf}-\frac{2\nu}{W}\ket{ee}-\ket{fg}+\ket{ff},\nonumber\\
\ket{S_{\pm}}&=\ket{gg}\BT{+}\frac{W^2}{W^2+2\nu^2\pm2\nu\sqrt{W^2+\nu^2}}\ket{gf}\nonumber\\
&-\frac{2\left(\mp\nu+\sqrt{W^2+\nu^2}\right)}{W}\ket{ee}\nonumber\\
&+\frac{W^2}{W^2+2\nu^2\pm2\nu\sqrt{W^2+\nu^2}}\ket{fg}+\ket{ff}.\nonumber\\
\end{align}
Under the assumption of $\nu\sim W$, the stray eigenstates stay far from the logical space in energy. The on-resonance QR sidebands continuously pump the error states after single-photon loss to the target logical states, with an extra photon excitation appearing in the corresponding resonator $R_{j}$. These excitations in the resonators decay quickly at a rate $\kappa$ and recover the state. 

From another view, the detuned QQ red and blue sideband pairs are topologically equivalent to a four-pointed star in the sideband configuration with $\ket{ee}$ as the center, shown in Fig.~\ref{fig:energy_level}(b). This effectively introduces the 4-photon sidebands $\ket{gf}\leftrightarrow\bra{fg}$ and $\ket{gg}\leftrightarrow\bra{ff}$ to the system, with $\ket{L_0}$ and $\ket{L_1}$ being separately the dark state of each 4-photon sideband. Since all other states are separated from the logical manifold by $O \left ( W \right )$ energy differences, the four QQ sidebands induce a dynamical decoupling effect that suppresses dephasing from low-frequency phase noise, just as in the original VSLQ proposal. The other bright states' eigenenergies are separated from the codewords through the QQ sideband frequency detuning $\{\nu_{0},\nu_{1}\}$, so that passive error correction does not mix the error states with them. Since $\ket{ee}$ is orthogonal to the dark states, the logical manifold stabilization happens through independent paths. Notice that the superposition state $\ket{L_x}=\ket{L_0}+\ket{L_1}$ in the frame of Eq.~\ref{eq:fully_rotated} will have a fast oscillating phase between logical basis. The energy shift to codewords comes simply from rotating frame choices and has no physical consequence.

{The Star Code also suppresses the no-jump error as the always-on two-qubit Hamiltonian $H_{QQ}$ maintains the logical state form. The suppression follows the same way as the suppression of 1/f dephasing noise, achieved through the dynamical decoupling effect.}

\section{AQEC Performance}

\begin{figure}[t]
    \centering
    \includegraphics[width=\columnwidth]{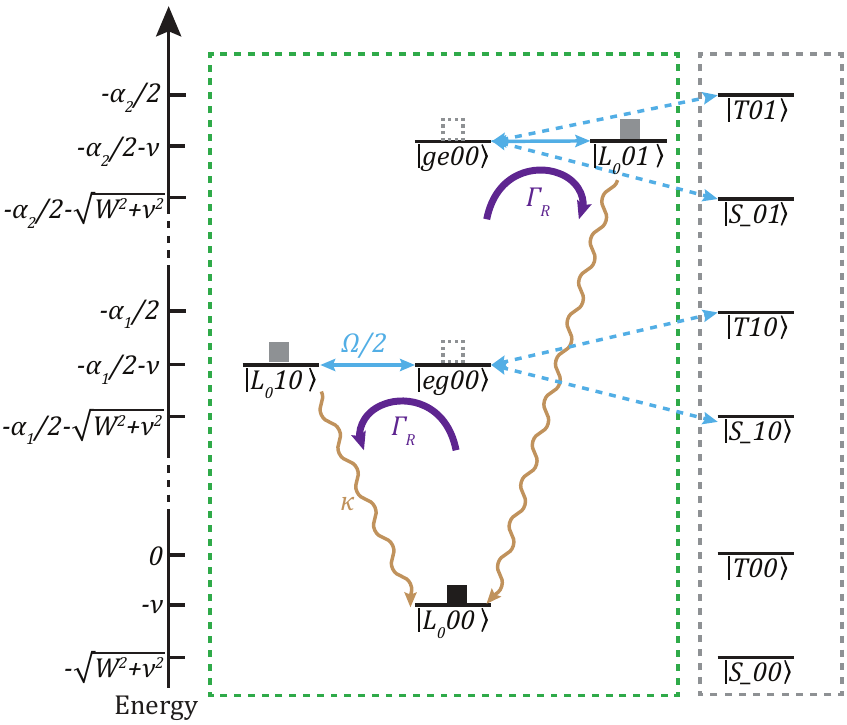}
    \caption{Error correction cycle for $\ket{L_0}$. The effective $\ket{L_0}$ refilling rate $\Gamma_{R}$ is shown in the purple arrow. A second photon loss can happen at rate $\gamma$ before the completion of the refilling cycle. Population transfer to the grey dash box is marked with blue dash arrows, and the population transfer is suppressed by the energy difference $O(\nu)$. }
    \centering
    \label{fig:analytic_lifetime}
\end{figure}

Next, we discuss the theoretical lifetime improvement against single-photon loss. We approximate the lifetime improvement semi-classically and verify its agreement using simulation. First, we consider the case of $\ket{L_0}$ and ignore the population lost to the stray eigenstates under the QR sideband. The logical states refilling rate $\Gamma_{R}$ is a two-step process: the QR sidebands that resonantly bring error states to the parent logical states, and the resonator photon loss. Using Fermi's golden rule and assuming Lorentzian distribution of lossy resonators' energy~\cite{eliot2014, kapit2017review}, we have $\Gamma_{R}=\frac{\Omega^{2}\kappa}{\kappa^{2}+\Omega^{2}}$. We label the population of $\ket{L_000}$ and $\ket{eg00}$ (also for $\ket{ge00}$) at time $t$ as $P_L(t)$ and $P_E(t)$. {By symmetry, the population of $\ket{eg00}$ and $\ket{ge00}$ are the same} Assuming the system started with $\ket{L_000}$ at time $t=0$, we can express the evolution using the following differential equations
\begin{equation}
    \begin{cases}
      \frac{dP_{L}(t)}{dt}= & -2\gamma P_{L}(t)+2\Gamma_{R}P_{E}(t)\\
      \frac{dP_{E}(t)}{dt}= & \gamma P_{L}(t)-\left(\gamma+\Gamma_{R}\right)P_{E}(t)\\
      P_{L}(0)= & 1\\
      P_{E}(0)= & 0
    \end{cases}.       
\end{equation}
The solution of $P_{L}(t)$ has two parts, a fast exponential decay term with a small weight, and a dominant slow exponential decay term. Assuming $\Gamma_{R}\gg\gamma$, the slow decay term shows quadratic lifetime improvement, compared to the physical transmon decay rate $\gamma$.
\begin{equation}
\begin{cases}
    P_L(t) = & \frac{-\gamma+\Gamma_R+\Delta}{2\Delta}\exp\left(t(\Delta-3\gamma-\Gamma_R)/2\right)\\
    &+\frac{\gamma-\Gamma_R+\Delta}{2\Delta}\exp\left(t(-\Delta-3\gamma-\Gamma_R)/2\right) \\
    &\approx 
    (1-2\gamma/\Gamma_R)\exp\left(-\frac{2\gamma^2t}{\Gamma_{R}+3\gamma}\right)
    \label{eq:the_improve}\\
    \Delta = & \sqrt{\gamma^2+6\gamma\Gamma_R+\Gamma_R^2}
\end{cases}
\end{equation}

Now we introduce the stray eigenstates $\{\ket{S_-},\ket{T},\ket{S_+}\}$ into the system. As shown in Fig.~\ref{fig:analytic_lifetime}, the population transfer from the error states to the stray eigenstates is also a two-step process. By keeping the closest two eigenstates $\ket{S_{-}}$ and $\ket{T}$ in terms of energy, the refilling rates $\{\Gamma_{S},\Gamma_{T}\}$ are
\begin{equation}
    \begin{cases}
      \Gamma_{S}= & \frac{\kappa\Omega^2k_s}{4\left(-\nu+\sqrt{W^2+\nu^2}\right)^2+\kappa^2+\Omega^2k_s}\\      
      \Gamma_{T}= & \frac{\kappa\Omega^2/\left(1+\frac{\nu^2}{W^2}\right)}{16\nu^2+4\kappa^2+\Omega^2/\left(1+\frac{\nu^2}{W^2}\right)}  \\ 
      k_s = & (\bra{S_{-}}\ket{fg})^2
    \end{cases}.     
    \label{eq:L0 det}
\end{equation}
Again assuming the initial state at the beginning to be $\ket{L_000}$ and treating population to $\{\ket{S_{-}},\ket{T}\}$ as an uncorrectable logical coherence loss, we have the following equations of motion
\begin{equation}
    \begin{cases}
      \frac{dP_{L}(t)}{dt}= & -2\gamma P_{L}(t)+2\Gamma_{R}P_{E}(t)\\
      \frac{dP_{E}(t)}{dt}= & \gamma P_{L}(t)-\left(\gamma+\Gamma_{R}+\Gamma_{S}+\Gamma_{T}\right)P_{E}(t)\\
      P_{L}(0)= & 1\\
      P_{E}(0)= & 0
    \end{cases}.     
\end{equation}
\begin{figure}[t]
    \centering
    \includegraphics[width=\columnwidth]{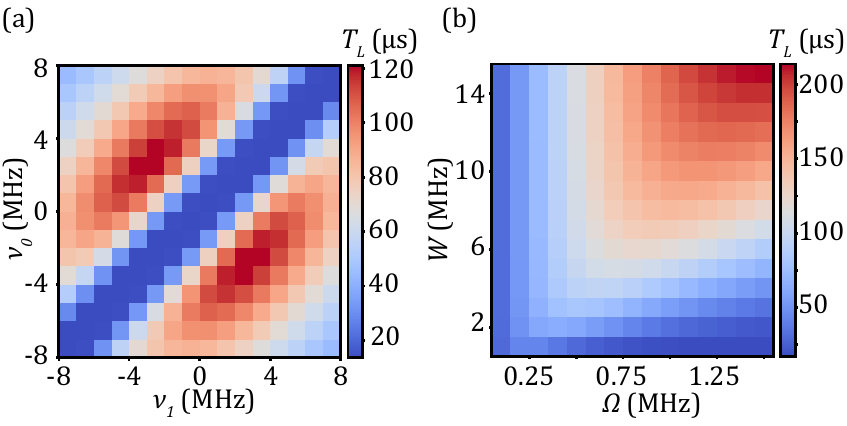}
    \caption{Logical lifetime ($T_L$) as a function of detunings and sideband rates. Simulations are performed up to $\SI{200}{\micro\second}$ with $T_1=\SI{20}{\micro\second}$ for both transmons. The logical $T_L$ are extracted by fitting the last $\SI{180}{\micro\second}$ to an exponential decay profile. (a) 2D scan of QQ sideband detunings $\nu_{0}$ and $\nu_1$. Other parameters used in the simulation: $\{\alpha_{1},\alpha_{2},W,\Omega,\kappa\}=\{-160,-260,5,1,0.5\}$MHz. Optimal performance is obtained around $\nu_0=-\nu_1=\pm W/\sqrt{3}$. (b) 2D scan of QQ and QR sideband rates $W$ and $\Omega$. Parameters are set to be $\nu_0=-\nu_1=W/\sqrt{3}$, and $\Omega=\kappa$ for best AQEC performance. Simulations show significantly improved performance around $\Omega=W/10$.}
    \centering
    \label{fig:2d_det_sweep}
\end{figure}
Given $\Gamma_{R}\gg\gamma,\Gamma_{S}, \Gamma_{T}$, the slow decay rate in $P_{L}(t)$ is
\begin{equation}
\Gamma_{L0}\sim\frac{2\gamma(\gamma+\Gamma_{S}+\Gamma_{T})}{3\gamma+\Gamma_{R}+\Gamma_{S}+\Gamma_{T}}.
\label{eq:L0 lifetime}
\end{equation}
The slow decay rate for $\ket{L_1}$ can be derived similarly
\begin{equation}
\Gamma_{L1}\sim\frac{2\gamma(3\gamma+\Gamma_{S}+\Gamma_{T})}{5\gamma+\Gamma_{R}+\Gamma_{S}+\Gamma_{T}}.
\label{eq:L1 lifetime}
\end{equation}
Note that, for the realistic parameter ranges considered in this work, $\Gamma_S$ and $\Gamma_T$ will be much smaller than $\gamma$ and are effectively negligible in determining the logical decay rates.

Using Eq.~\ref{eq:L0 det}, \ref{eq:L0 lifetime} and \ref{eq:L1 lifetime} one can verify that larger QQ sideband rate $W$ and detunings $\nu$ will provide better energy isolation, leading to a higher logical states' lifetime. The ratio  $\Gamma_{L1}/\Gamma_{L0} \sim 3$ indicates that the logical qubit has approximately $3$ times faster decay rate than the excitation rate, as the average photon number of $a\ket{L_1}$ (error state) is three times larger than that of $a\ket{L_0}$. Depolarization rate $\Gamma_{Z}$ for the logical state is $\Gamma_{Z}=\Gamma_{L0}+\Gamma_{L1}$. For the transversal dephasing rate $\Gamma_{X}$, extra protection comes from the code structure. When a double-photon loss event happens on the same physical qubit (with $50\%$ chance), the state obtains $50\%$ overlap with $\ket{L_x}$. Therefore, for a quarter of the double-photon loss event, $\ket{L_x}$ does not experience coherence loss, and the lifetime for $\ket{L_x}$ is $T_{X}=4T_{Z}/3$. For both $T_{Z}$ and $T_{X}$, the lifetime improvement is quadratic given $\Gamma_{S}+\Gamma_{T}\ll\gamma$.

\begin{figure}[t]
    \centering
    \includegraphics[width=\columnwidth]{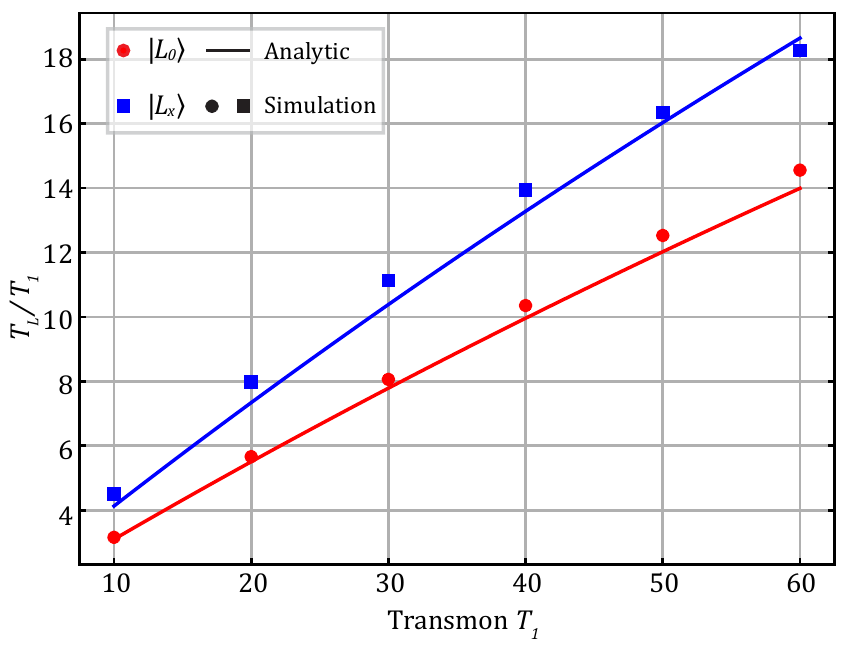}
    \caption{Logical lifetime improvement as a function of transmon $T_{1}$ (considered identical for both transmons). Quadratic lifetime improvement (roughly linear improvement in the lifetime ratio) under AQEC is clearly seen in the plot. Logical $T_{L}$ are extracted by fitting traces to the exponential decay curve $A\exp(-t/T_L)+C$ (with $A$ and $C$ being free parameters), and the improvement ratio is $T_L/T_{1}$. Error bars (one standard deviation) for $T_L$ are smaller than the marker size. Each simulation is run up to $\SI{800}{\micro\second}$, and the short period is not included in the fitting. Other parameters used in the simulation are $\{\alpha_1, \alpha_2, W, \nu_0, \nu_1, \Omega, \kappa\}=\{-160, -260, 10,5.77,-5.77,0.71,0.5\}$MHz. The analytic expression (solid lines) matches the simulation result. The depolarization lifetime of $\ket{L_1}$ is almost the same as $\ket{L_0}$ in simulation. All simulated logical lifetimes here are above the break-even point.}
    \centering
    \label{fig:T1_sweep}
\end{figure}

We perform rotating-frame simulations to verify the lifetime improvements. Fig.~\ref{fig:2d_det_sweep}(a) shows the lifetime of $\ket{L_x}$ under different QQ sideband detuning combinations. We neglect short timescale behavior when extracting logical states' lifetimes. There is a low coherence strip along the diagonal region. This happens when $\nu_{0}=\nu_{1}$, as $\{\ket{L_0},\ket{L_1},\ket{gg}+\ket{ff}-\ket{gf}-\ket{fg}\}$ become degenerate eigenstates with non-orthogonal error states and violates the Knill-Laflamme condition. From Fig.~\ref{fig:2d_det_sweep}(a), the maximum lifetime improvement region appears around $\nu_{0}=-\nu_{1}=\pm W/\sqrt{3}$. This can be intuitively understood as $\{\ket{S_-},\ket{L_0},\ket{T},\ket{L_1},\ket{S_+}\}$ are evenly separated in energy (Fig.~\ref{fig:energy_level}(c)), thus providing close-to-optimal suppression of leakage to non-logical state population.

We fix the detuning relation $\nu_{0}=-\nu_{1}=W/\sqrt{3}$ and sweep $W, \Omega$ for $\ket{L_x}$'s lifetime. The results are plotted in Fig.~\ref{fig:2d_det_sweep}(b). During the sweep, we choose $\kappa=\Omega$, where refilling rate $\Gamma_{R}$ are optimal and error correction performance becomes insensitive to small changes in $\kappa$. In practice, $W=\SI{10}{\mega\hertz}$ and $\Omega=\SI{1}{\mega\hertz}$ can be achieved in current devices with some optimization~\cite{luchakram2017, Brown2022, li2023autonomous}. Since larger $W$ is more difficult to achieve in the system, given maximum $W$, optimal performance appears along the diagonals, where $\Omega$ is roughly an order of magnitude smaller than $W$. Finally, we sweep $T_{1}$ of the transmons and show the ratio of logical to physical lifetime in Fig.~\ref{fig:T1_sweep}. The quadratic improvement in logical states' lifetime is clearly visible and the data match pretty well with the analytic expression.

We note that the logical lifetime limit from other error channels (e.g. $1/f$ noise-induced dephasing and comparatively rare random photon addition due to finite temperature) in the Star code protocol is the same as in Ref.~\onlinecite{eliot2016}, because the AQEC process is the same except for a different Hamiltonian construction. To prevent leakage to higher transmon energy levels, the two transmons are chosen to have large but different anharmonicities $\alpha_{j}$. This will highly suppress blue sideband transitions such as $\ket{gf}\bra{eh}$, and red sideband transitions such as $\ket{fe}\bra{hg}$ that populates $\ket{h}$ level. These leakages are essentially negligible for the range of parameters considered in simulations.

\section{Gate protocols and discussion}

We now discuss logical operations on the Star code device. To be fully useful for quantum computing, any small logical qubit design should reduce gate error, in addition to extending idle lifetime. This almost certainly requires error transparency~\cite{eliot2018}, where gate waveforms are carefully tuned such that the gate Hamiltonian commutes with a single photon loss operator, when acting on the logical state manifold. Since the Star code uses the same code structure as the VSLQ, the error-transparent gate set introduced for the VSLQ can be directly generalized to it with minor modifications. {In particular, since they rely on dynamically generated dispersive shift terms, the error transparent $Z_L=\left(\ket{f}\bra{f}+\ket{e}\bra{e}-\ket{g}\bra{g}\right)\left(\ket{f}\bra{f}+\ket{e}\bra{e}-\ket{g}\bra{g}\right)$ and two-logical-qubit $CZ$ gates $Z_{L1}Z_{L2}$ can be implemented with linear couplers, like the star code itself. Here $Z_{L1}$ and $Z_{L2}$ are separately the error transparent $Z_L$ operator for each Star Code unit. Realizing error transparent CZ gate requires perturbative engineering of ZZ
coupling in both logical units and remains future simplification.} The single-logical-qubit error transparent $X_L$ operator, however, requires three-photon processes at the minimum and thus is much more difficult to engineer with a linear coupler. One could however implement it using a nonlinear coupler such as a SNAIL (Superconducting Nonlinear Asymmetric Inductive eLements)~\cite{snail2017} instead of the inductive shunt used in current experiments~\cite{li2023autonomous}, or find alternative ways to generate it not explored in the original error transparency work. 

One could also implement the Star code using more complex objects as the base qubit. Virtually any superconducting qubit design can be used in place of the transmons considered here, provided that it has three workable energy levels, significant nonlinearity, an error structure such that $\ket{f}$ decays directly to $\ket{e}$ with no single photon coupling to $\ket{g}$, and is compatible with AC-driven tunable couplers. A particularly interesting possibility would be to generalize the Star code to \textit{linear} objects such as coupled 3D cavities, given the substantially higher base coherence such devices exhibit compared to planar circuits. This is hardly a trivial enterprise given that, for example, one can no longer selectively drive $\ket{ee} \to \ket{gf}$ without also resonantly driving $\ket{ge} \to \ket{eg}$ in such a system, but we expect a suitably clever generalization of the Hilbert space topology and dark state structure of the Star code could be possible for linear systems as well. Of course, if the cavities have Kerr nonlinearities (due to interactions with transmon qubits or similar) then the generalization of the Star code to them is much simpler, though these nonlinearities are typically at least two orders of magnitude smaller than in a transmon. Such extensions could be a fruitful line of future research.

{In summary, we provided the optimal parameter choices for a novel error correction code called the Star code, that can correct single photon losses and suppress dephasing fully autonomously using just two tunably coupled qubits and two resonators.} It originates from the earlier VSLQ protocol but is substantially simpler to implement as it requires only two-photon interactions. It is capable of achieving quadratic lifetime improvements in the logical state lifetime by carefully tuning the circuit and drive parameters. As tunable couplers have become an increasingly popular route to the high-fidelity operation of multi-qubit circuits, this greatly simplified logical qubit design can be readily implemented in many existing superconducting qubit platforms.

\section{ACKNOWLEDGEMENTS}
We would like to thank Srivatsan Chakram and Nick Materise for their useful discussions in formulating this code. This work was supported by AFOSR Grant No.
FA9550-19-1-0399 and ARO Grant No. W911NF-17-S0001. Devices are fabricated in the Pritzker Nanofabrication Facility at the University of Chicago, which receives
support from Soft and Hybrid Nanotechnology Experimental (SHyNE) Resource (NSF ECCS-1542205), a node
of the National Science Foundation’s National Nanotechnology Coordinated Infrastructure. This work also made use of the shared facilities at the University of Chicago Materials Research Science and Engineering Center, supported by the National Science Foundation under award number DMR-2011854. EK's research was additionally supported by NSF Grant No. PHY-1653820.

\pagebreak

\appendix

\section{Logical lifetime limits from non-ideal parameters}

Here, we discuss the logical lifetime limits from non-ideal parameters. 

\begin{figure}[t]
    \centering
    \includegraphics[width=\columnwidth]{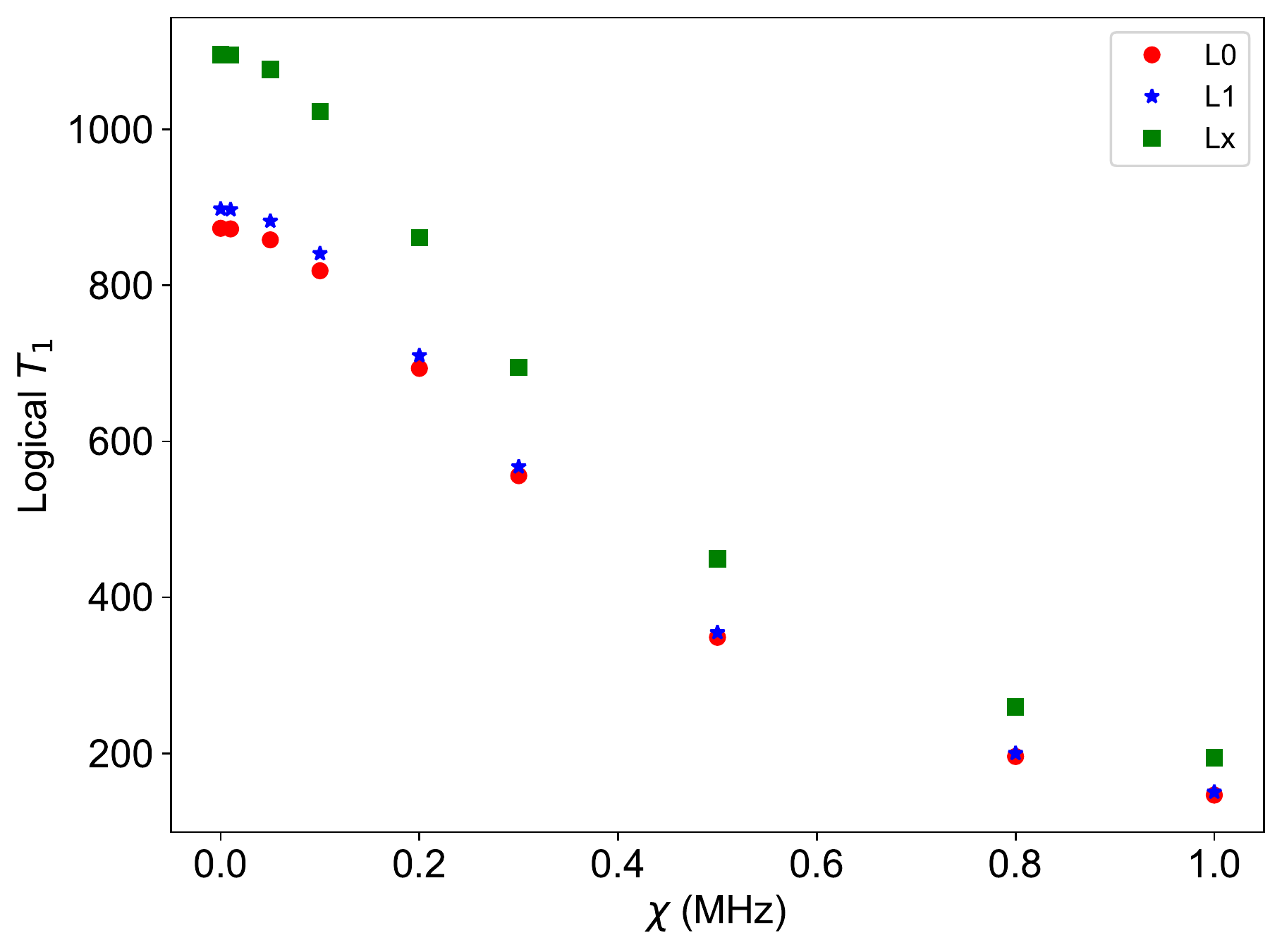}
    \caption{Simulated logical states' lifetime in the presence of qubit-resonator coupling $\chi$. Parameter used in the simulation: $\{\alpha_1, \alpha_2, W, \nu_0, \nu_1, \Omega, \kappa\}=\{-160, -260, 10,5.77,-5.77,0.71,0.5\}$ MHz, $T_1=\SI{60}{\micro\second}$.}
    \centering
    \label{fig:QR_ZZ}
\end{figure}

The ZZ interactions between transmons (Q) and readout resonators (R) are needed to distinguish the transmon state. While Star Code only requires the XX interactions between QR, the presence of QR dispersive coupling $\chi$ helps to calibrate the system. Fig.~\ref{fig:QR_ZZ} shows the logical state lifetime master equation simulation for $\ket{L_0}$, $\ket{L_1}$, and $\ket{L_x}$
in the presence of QR ZZ coupling. The logical state lifetime is weakly reduced in the low $\chi$ regime. This is because the photon decay from either resonator will have different frequencies depending on the coupled transmon being $\ket{g}$ or $\ket{f}$. Such a resonator-induced dephasing does not introduce a logical dephasing error but only distorts the form of $\ket{L_0}$ and $\ket{L_1}$. This noise has a Lorentzian spectrum that decays in frequency. Once the QQ sideband rate $W$ is much larger than $\chi$, the resonator-induced dephasing will be suppressed strongly as the 1/f dephasing noise. Therefore, in Fig.~\ref{fig:QR_ZZ} the logical $T_1$ is insensitive to the presence of small $\chi$.

Photon excitation in the readout resonators is detrimental to the Star Code. Suppose $R_1$ excites a photon when the logical state is $\ket{L_0}$, The QR sideband $\ket{L_010}\leftrightarrow\ket{eg00}$ will be activated and convert the logical state into the error state. This becomes a potential logical error unless the error state is flipped back before the second photon loss error happens.

The ZZ interactions between two transmons dephase the logical superposition state. Among all the ZZs between two qutrits, $ZZ_{ff1}=E_{\ket{ff}}-E_{\ket{ef}}-(E_{\ket{fg}}-E_{\ket{eg}})$ and $ZZ_{ff2}=E_{\ket{ff}}-E_{\ket{ef}}-(E_{\ket{gf}}-E_{\ket{ge}})$ will cause the logical state dephasing, as a random phase between $\ket{L_0}$ and $\ket{L_1}$ will accumulate, which is proportional to the product of time error is corrected and $ZZ_{ffi}$. Longer transmon $T_1$ and faster error correction rate (increasing QR sideband rate $\Omega$) help mitigate such dephasing channel, and the cancellation requires a simultaneous zero $ZZ_{ff1}$ and $ZZ_{ff2}$ when all QQ sidebands are on. This is achievable by adding extra detuned drives, such as the scheme discussed in Ref.~\cite{ZZcancelation1} and Ref.~\cite{ZZcancelation2}.

The Star Code is insensitive to the fluctuation in the QR sideband rate $\Omega_i$ and does not require $\Omega_1=\Omega_2$ (Used only for main text equation simplicity). Fluctuations in both QQ sideband rate $W$ and detunings $\nu_i$ are strongly suppressed as long as they are not comparable to the energy gap ($O(W)$) between $\ket{L_0}$ and $\ket{L_1}$.

\section{Logical state process fidelity}
\begin{figure}[t]
    \centering
    \includegraphics[width=\columnwidth]{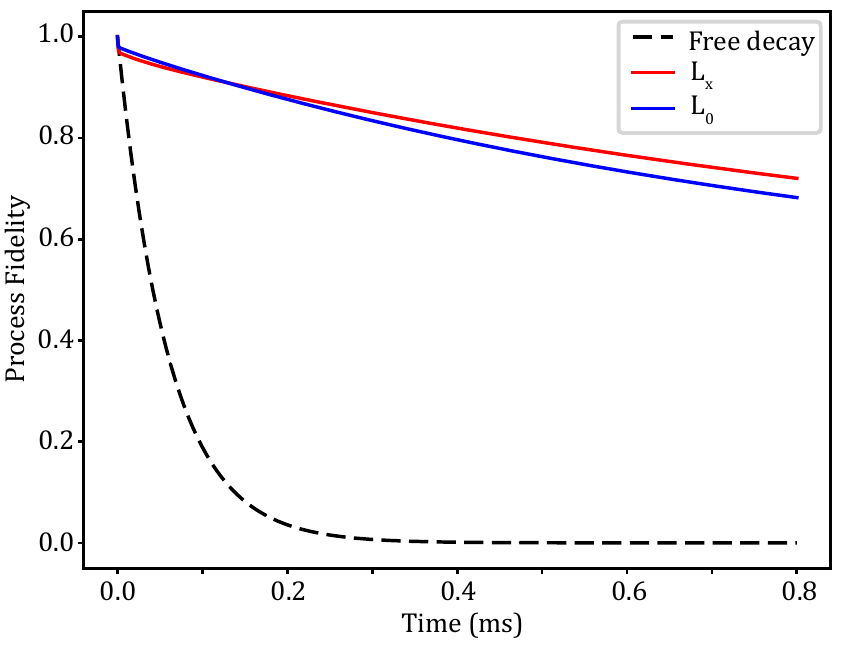}
    \caption{Simulated logical states' lifetime and physical qubit lifetime. Parameter used in the simulation: $\{\alpha_1, \alpha_2, W, \nu_0, \nu_1, \Omega, \kappa\}=\{-160, -260, 10,5.77,-5.77,0.71,0.5\}$ MHz, $T_1=\SI{60}{\micro\second}$.}
    \centering
    \label{fig:process_fidelity}
\end{figure}

Fig.~\ref{fig:process_fidelity} shows the simulated process fidelity for $\ket{L_x}$, $\ket{L_0}$, and physical qubit decay. Operators used for calculating the process fidelity for $\{\ket{L_0}, \ket{L_x}\}$ are $\{\ket{L_0}\bra{L_0}-\ket{L_1}\bra{L_1}, \ket{L_x}\bra{L_x}\}$. 

\bibliography{2Q}
\end{document}